\documentclass[prb,twocolumn,showpacs]{revtex4}
\usepackage{amsmath}
\usepackage{dcolumn}
\usepackage{graphicx}
\usepackage{bm}
\usepackage{dsfont}
\usepackage{hyperref}

\hypersetup{
     citecolor = {green},
   colorlinks = {true},                      urlcolor = {blue} }

\begin{document}

\title{Intermediate Haldane phase in spin-2 quantum chains with uniaxial
anisotropy}
\author{Hong-Hao Tu and Rom\'an Or\'us}
\affiliation{Max-Planck-Institut f\"ur Quantenoptik, Hans-Kopfermann-Stra\ss e 1, 85748
Garching, Germany}
\date{\today}

\begin{abstract}
We provide evidence of an intermediate Haldane phase in a spin-2 quantum
chain. By combining effective field theory and numerical approaches, we show
that the phase diagram of the proposed model includes SO(5) Haldane,
intermediate Haldane, and large-$D$ phases. We determine the characteristic
properties of these phases, including edge states, string order parameters,
and degeneracies in the entanglement spectrum. The symmetries responsible
for the degeneracy patterns observed in the entanglement spectrum are also
discussed.
\end{abstract}

\pacs{75.10.Pq, 75.10.Jm, 03.67.Mn}
\author{}
\maketitle

\textit{Introduction.} Characterization of quantum phases beyond Landau's
symmetry breaking paradigm \cite{Landau-1937} is an important open problem
in physics. Recently, for one-dimensional (1D) gapped phases, a
classification scheme based on matrix product states (MPS) has been put
forward \cite{Gu-2009,Pollman-2009,Schuch-2010}, based on the fact that
ground states of 1D gapped Hamiltonians can be efficiently approximated by
MPS \cite{Verstraete-2006,Hastings-2007}. This scheme complements the
conventional approach that classifies relevant perturbations of fixed
points, and sheds new light on some phases that have been extensively
studied, e.g., the Haldane phase in integer-spin chains \cite{Haldane-1983}.

These developments may be helpful in order to clarify a controversial
problem: the possibility of an intermediate Haldane (IH) phase (also called
intermediate-$D$ phase) in quantum spin chains. This problem originated from
the study of a spin-2 Heisenberg chain with uniaxial anisotropy $H=$ $%
\sum_{j}\vec{S}_{j}\cdot \vec{S}_{j+1}+D\sum_{j}(S_{j}^{z})^{2}$, where $%
D\geq 0$ (here $\vec{S}$ and $S^{z}$ are the usual spin operators). For $D=0$%
, the ground state is in the so-called Haldane phase. For $D\rightarrow
\infty $, the ground state is in a large-$D$ phase, close to a trivial
product state with $S_{j}^{z}=0$ for all $j$. Regarding the phase diagram of
this model, a field theoretical approach \cite{Schulz-1986}\ suggests a
single phase transition between the Haldane and the large-$D$ phases. On the
contrary, for intermediate $D$, Oshikawa \cite{Oshikawa-1992} predicted\
that an IH phase may emerge between the two phases. He justified this by
noticing that $S_{j}^{z}=\pm 2$ states are substantially suppressed by the $D
$ term but $S_{j}^{z}=\pm 1$ states are less affected, which may lead to the
formation of an effective spin-1-like Haldane phase with residual $%
S_{j}^{z}=\pm 1,0$ states. Nevertheless, whether such an IH phase really
exists or not in the above spin-2 model (and its generalizations) remains
unclear \cite%
{Oshikawa-1995,Schollwock-1996,Nomura-1998,Hatsugai-2008,Tonegawa-2011}.

In this work, we study a spin-2 quantum chain for which we provide sharp
evidence of the existence of an IH phase. As far as we know, our results
provide the clearest evidence so far in favor of the existence of such a
phase in quantum spin chains. Our search for the IH phase is guided by an
effective field theory, which yields a qualitative phase diagram with three
quantum phases, namely: SO(5) Haldane, IH, and large-$D$ phases. The field
theory also determines characteristic features of the phases, e.g., edge
states in open chains and string order parameters (SOPs) \cite{den Nijs-1989}%
. We also determine numerically the phase diagram of the spin model, and
find full agreement with the field-theory predictions. Moreover, we study
the entanglement spectrum (ES) \cite{Haldane-2008} of the different phases,
and confirm that the IH phase has a \textit{double} degeneracy in the ES
that is protected by symmetries, which distinguishes itself from the SO(5)
Haldane phase with \textit{quadruple} degeneracy and the large-$D$ phase
without protected degeneracy. The spatial inversion, time reversal, and $%
(Z_{2}\times Z_{2})^{2}$ symmetries responsible for the robust degeneracy of
the ES are also investigated.

\textit{Model Hamiltonian and symmetries.} In this work, we consider the
spin-2 quantum chain
\begin{equation}
H=\sum_{j}\sum_{\gamma =1}^{4}J_{\gamma }(\vec{S}_{j}\cdot \vec{S}%
_{j+1})^{\gamma }+D\sum_{j}(S_{j}^{z})^{2},  \label{eq:Hamiltonian}
\end{equation}
for periodic boundary condition and in the thermodynamic limit. In our case,
we take $J_{1}=-\frac{11}{6},J_{2}=-\frac{31}{180},J_{3}=\frac{11}{90},J_{4}=%
\frac{1}{60}$, and consider the region $D\geq 0$. As we shall see, this
generalization of the usual spin-2 Heisenberg chain has a number of
important properties.

First, let us identify the symmetries of Eq. (\ref{eq:Hamiltonian}), which
will turn out to be very useful for our purposes. For $D=0$, the Hamiltonian
in Eq. (\ref{eq:Hamiltonian}) can be rewritten as $H_{D=0}=2%
\sum_{j}[P_{2}(j,j+1)+P_{4}(j,j+1)]$, where $P_{S_{T}}(j,j+1)$ projects onto
total spin-$S_{T}$ states of neighboring sites $j$ and $j+1$. This model has
SO(5) symmetry and an MPS as its exact ground state \cite%
{Tu-2008,Scalapino-1998}. To identify the SO(5) symmetry, we work in the
standard $S^{z}$ basis $|m\rangle $ ($m=\pm 2,\pm 1,0$)\ and define SO(5)
Cartan generators $L^{12}=|2\rangle \langle 2|-|-2\rangle \langle -2|$ and $%
L^{34}=|1\rangle \langle 1|-|-1\rangle \langle -1|$. By defining $L^{15}=%
\frac{1}{\sqrt{2}}(|2\rangle \langle 0|+|0\rangle \langle -2|+\mathrm{h.c.})$
and $L^{35}=\frac{1}{\sqrt{2}}(|1\rangle \langle 0|+|0\rangle \langle -1|+%
\mathrm{h.c.})$, the SO(5) commutation relations $[L^{ab},L^{cd}]=i\left(
\delta _{ac}L^{bd}+\delta _{bd}L^{ac}-\delta _{ad}L^{bc}-\delta
_{bc}L^{ad}\right) $ fix the ten generators $L^{ab}$ $(1\leq a<b\leq 5)$.
For $D=0$, the Hamiltonian commutes with all ten operators $%
\sum_{j}L_{j}^{ab}$ and therefore has SO(5) symmetry.

For $D>0$, this SO(5) symmetry is explicitly broken down to U(1)$\times $%
U(1). In order to see this, we rewrite the uniaxial anisotropy as $%
(S^{z})^{2}=4(L^{12})^{2}+(L^{34})^{2}$. Thus, $\sum_{j}L_{j}^{12}$ and $%
\sum_{j}L_{j}^{34}$ commute not only with each other but also with $H$, and
therefore the model has U(1)$\times $U(1) symmetry. Additionally, the Hamiltonian
in Eq.(\ref{eq:Hamiltonian}) also has discrete symmetries, including spatial
inversion, time reversal, and a set of $Z_{2}$ symmetries. These $Z_{2}$
symmetries are a consequence of the invariance under global $Z_{2}$
rotations $U^{ab}=e^{i\pi L^{ab}}$ for all $L^{ab}$. The $Z_{2}$ operators
form a $(Z_{2}\times Z_{2})^{2}$ group, whose elements, without loss of
generality, can be chosen as $\{\mathds{1},U^{12}\}\times \{\mathds{1}%
,U^{15}\} \times \{\mathds{1} ,U^{34}\}\times \{\mathds{1},U^{35}\}$. Here
we remind that these $Z_{2}$ operators preserve their form under the
nonlocal Kennedy-Tasaki transformation \cite{Kennedy-1991} and also generate
a hidden $(Z_{2}\times Z_{2})^{2}$ symmetry in dual space \cite{Tu-2008}.

\textit{Field-theory treatment.} Even though the model in Eq. (\ref%
{eq:Hamiltonian}) looks quite complicated, its effective field theory at
low energy is very simple. As we shall explain soon, this is given by the
following Hamiltonian density of five Majorana fermions $\xi ^{a}$ ($%
a=1,\ldots ,5$):%
\begin{eqnarray}
\mathcal{H}_{\mathrm{eff}} &=&-iv\sum_{a=1}^{5}(\xi _{R}^{a}\partial _{x}\xi
_{R}^{a}-\xi _{L}^{a}\partial _{x}\xi _{L}^{a})-im_{1}\sum_{a=1}^{2}\xi
_{R}^{a}\xi _{L}^{a}  \notag \\
&&-im_{2}\sum_{a=3}^{4}\xi _{R}^{a}\xi _{L}^{a}-im_{3}\xi _{R}^{5}\xi
_{L}^{5},  \label{eq:FT}
\end{eqnarray}%
where $v$ and $m_{a}$ are velocity and masses of the Majorana fermions, and
marginal four-fermion interactions have been neglected.

The strategy to derive Eq. (\ref{eq:FT}) is to start from the SO(5) point $D=0
$, whose effective field theory is known to be of the form given by Eq.(\ref%
{eq:FT}) with $m_{1}=m_{2}=m_{3}<0$ \cite{Alet-2010}. In the continuum
limit, the effect of the $D$ term can be taken into account by using
bosonization techniques. Following Ref. \cite{Alet-2010,Tu-Orus-2011}, we
find that $\sum_{j}(L_{j}^{12})^{2}\sim ig\int dx\sum_{a=1}^{2}\xi
_{R}^{a}\xi _{L}^{a}$ and $\sum_{j}(L_{j}^{34})^{2}\sim ig\int
dx\sum_{a=3}^{4}\xi _{R}^{a}\xi _{L}^{a}$, where $g<0$. By using $%
(S_{j}^{z})^{2}=$ $4(L_{j}^{12})^{2}+(L_{j}^{34})^{2}$, we arrive then at
the expression given in Eq. (\ref{eq:FT}). For $D\rightarrow 0$, we have $%
m_{1}-m_{3}\simeq 4(m_{2}-m_{3})\propto D$, but this relation does not hold
for larger $D$ due to renormalization effects. Thus, the Majorana masses\
and the velocity $v$\ are treated as phenomenological parameters. However,
the fact that only three independent masses appear in Eq. (\ref{eq:FT}),
which ensures O(2)$\times $O(2) symmetry, is imposed by the U(1)$\times $%
U(1)\ symmetry of Eq. (\ref{eq:Hamiltonian}) since O(2)$\simeq $U(1).
Moreover, the $(Z_{2}\times Z_{2})^{2}$ symmetry of Eq. (\ref{eq:Hamiltonian}%
) is also revealed by the invariance of Eq. (\ref{eq:FT})\ under $Z_{2}$
transformations $\xi _{R(L)}^{a}\rightarrow -\xi _{R(L)}^{a}$ \cite%
{Alet-2010}.

\begin{figure}[tbp]
\centering
\includegraphics[scale=0.36]{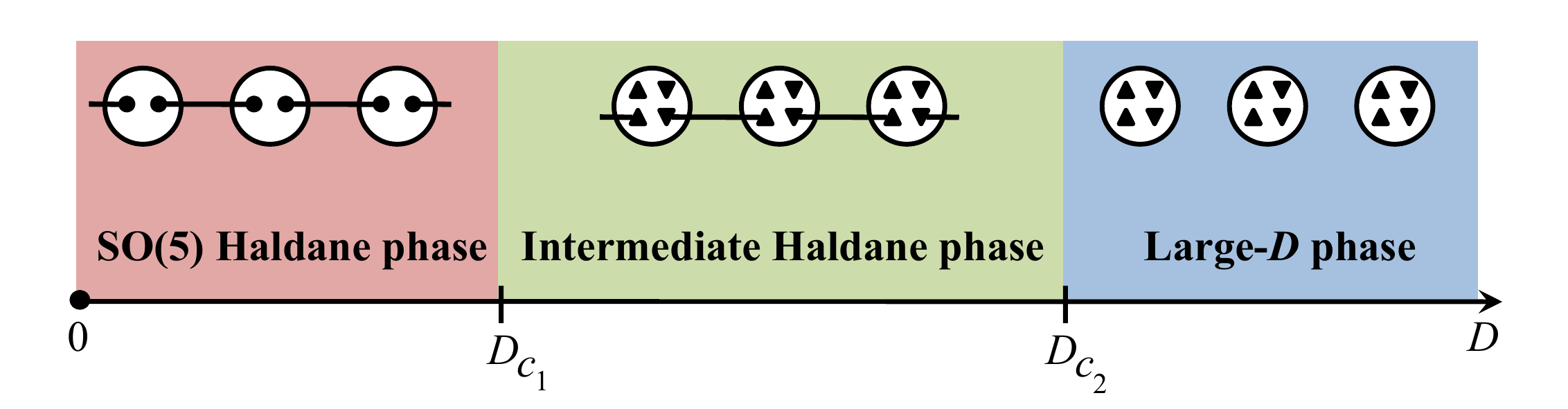}
\caption{(Color online) Qualitative phase diagram of the spin-2 model in Eq. (%
\protect\ref{eq:Hamiltonian}). The diagrams in the upper side correspond to
the MPS structure describing the renormalization group fixed point for each
phase: for low-$D$, ancillary spin-3/2 singlets are projected onto spin-2
subspaces at each physical site, thus providing an MPS for spin-2 physical
particles in terms of matrices of dimension 4; for intermediate-$D$,
ancillary spin-1/2 singlets, together with on-site spin-1/2 triplets $S^z=0$%
, are projected onto spin-2 at each site; for large-$D$, no singlets are
projected and the state is a product state.}
\label{fig:phasediagram}
\end{figure}

Armed with this effective field theory description, we are now in position
to sketch a phase diagram for the quantum spin chain in Eq. (\ref%
{eq:Hamiltonian}). When increasing $D$ from $0$ to $\infty $, we expect that
the Majorana mass $m_{3}$ is always negative in Eq. (\ref{eq:FT}), while $m_{1}
$ and $m_{2}$ change from negative to positive \textit{successively} at two
quantum critical points $D_{c_{1}}$ and $D_{c_{2}}$ ($D_{c_{1}}<D_{c_{2}}$).
Both critical theories at $D_{c_{1}}$ and $D_{c_{2}}\ $have two massless
Majoranas, and thus are equivalent to conformal field theories with central
charge $c=\frac{1}{2}\times 2=1$.\ For $0\leq D<D_{c_{1}}$, we call the
phase 'SO(5) Haldane phase', since its physics is captured by the SO(5)
point $D=0$. For $D_{c_{1}}<D<D_{c_{2}}$, the IH phase emerges, whose
characteristics will be discussed below. For $D>D_{c_{2}}$, the system
enters the large-$D$ phase. A qualitative phase diagram for (\ref%
{eq:Hamiltonian}) is shown in Fig. \ref{fig:phasediagram}.

Similar to the spin-1 Haldane phase \cite{Affleck-1987,TKNg-1994}, the
spin-2 IH phase has a bulk gap but exhibits gapless spin-1/2 edge
excitations in open chains. Let us have a closer look at how these edge
states emerge from Eq. (\ref{eq:FT}). For $m_{1}>0,m_{2}<0,m_{3}<0$, our
effective field theory description shares an analogy with Tsvelik's theory
\cite{Tsvelik-1990} for the spin-1 Haldane phase, formulated in terms of
three Majorana fermions. On a semi-infinite chain, both theories support
\textit{three} Majorana zero-energy modes at the boundary, forming a
fractionalized spin-1/2 edge state \cite{Lecheminant-2002}. This
distinguishes the IH phase from the SO(5) Haldane phase (with spin-3/2 edge
states \cite{Tu-2008} formed by five Majorana edge modes \cite{Alet-2010})
and the large-$D$ phase (without edge states). In fact, we expect that a
large family of spin-2 chains (namely, those described by a similar field
theory at low energy) also support an IH phase. Interestingly, the
corresponding Hamiltonians may capture the physics of some quasi-1D
compounds, and thus the fractionalized edge states in the IH phase may be
observed in electron spin resonance experiments by doping nonmagnetic ions
\cite{Hagiwara-1990}.

The effective field theory in Eq.(\ref{eq:FT}) also provides the order
parameters that characterize the three gapped phases. To see this, let us
view Eq.(\ref{eq:FT}) as five decoupled Ising models\ with the Majorana mass
$m\sim (T-T_{c})/T_{c}$, where $T_{c}$ is the Ising critical temperature
\cite{Alet-2010,Tu-Orus-2011}. Defining the Ising order and disorder
operators as $\sigma _{a}$ and $\mu _{a}$ $(a=1,\ldots ,5)$, the SO(5)
Haldane, IH, and large-$D$ phases correspond, respectively, to all five
Ising models, three Ising models ($a=3,4,5$), and just one Ising model ($a=5$%
)\ in the ordered phase(s) with $\langle \sigma _{a}\rangle \neq 0$. In
order to be able to distinguish these phases, we find that it is sufficient
to use two SOPs%
\begin{equation}
\mathcal{O}^{12}=\lim_{|k-j|\rightarrow \infty }\langle
L_{j}^{12}\prod_{l=j}^{k-1}\exp (i\pi L_{l}^{12})L_{k}^{12}\rangle
\label{eq:SOP}
\end{equation}%
and $\mathcal{O}^{34}$ (where $L^{12}$ is replaced by $L^{34}$). In the
continuum limit, these SOPs are related to Ising order operators as $%
\mathcal{O}^{12}\sim \langle \sigma _{1}\rangle \langle \sigma _{2}\rangle ,%
\mathcal{O}^{34}\sim \langle \sigma _{3}\rangle \langle \sigma _{4}\rangle $
\cite{Tu-Orus-2011}. Therefore, these SOPs distinguish the SO(5) Haldane ($%
\mathcal{O}^{12}\neq 0,\mathcal{O}^{34}\neq 0$), the IH ($\mathcal{O}^{12}=0,%
\mathcal{O}^{34}\neq 0$), and the large-$D$ ($\mathcal{O}^{12}=\mathcal{O}%
^{34}=0$) phases. As we shall see, this is very convenient in order to
evaluate numerically the phase diagram of the Hamiltonian in Eq. (\ref%
{eq:Hamiltonian}).

\begin{figure}[tbp]
\centering
\includegraphics[scale=0.17]{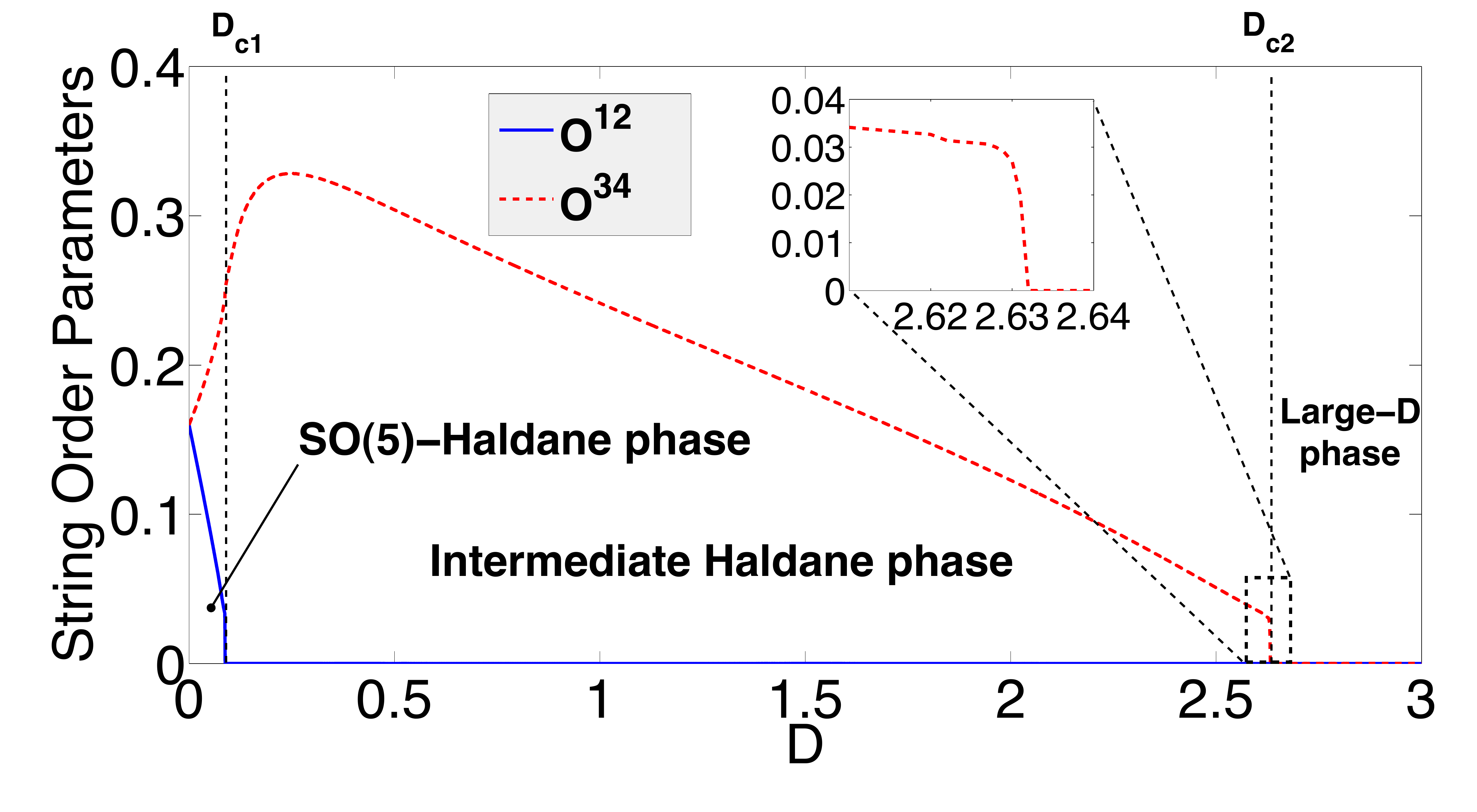}
\caption{(Color online) String order parameters $\mathcal{O}^{12}$ and $%
\mathcal{O}^{34}$. They vanish smoothly at $D=D_{c_1}$ and $%
D=D_{c_2}$, respectively.}
\label{fig:SOPs}
\end{figure}

The solvability of the SO(5) point $D=0$ provides an intuitive picture of
how these SOPs change with $D$. Starting from $D=0$, the MPS ground state of
Eq. (\ref{eq:Hamiltonian}) has a perfect hidden string order \cite{Tu-2008}:
In the $S^{z}$ basis, $\left\vert 2\right\rangle $ and $\left\vert
-2\right\rangle $ appear alternatively in all the configurations of the MPS
if $\left\vert 0\right\rangle $ and $\left\vert \pm 1\right\rangle $ are
removed. Similarly, $\left\vert 1\right\rangle $ and $\left\vert
-1\right\rangle $ also appear alternately, if $\left\vert 0\right\rangle $
and $\left\vert \pm 2\right\rangle $ are removed. This hidden string order
is reflected in a nonzero value of the SOPs, $\mathcal{O}^{12}=\mathcal{O}%
^{34}=0.16$. When increasing $D$, the uniaxial anisotropy in Eq. (\ref%
{eq:Hamiltonian}) tends to suppress both the $\left\vert \pm 2\right\rangle $
and $\left\vert \pm 1\right\rangle $ states, but with larger suppression
strength on $\left\vert \pm 2\right\rangle $. Thus, the string order for the
$\left\vert \pm 2\right\rangle $ states is destroyed earlier at $D_{c_{1}}$, 
and an IH phase is formed with remaining string order for the $\left\vert
\pm 1\right\rangle $ states. Here we remind that $L^{12}$ and $L^{34}$ act
on $\left\vert \pm 2\right\rangle $ and $\left\vert \pm 1\right\rangle $,
respectively. Therefore, we have $\mathcal{O}^{12}=0,\mathcal{O}^{34}\neq 0 $
in the IH phase. When increasing $D$ further, the string order for the $%
\left\vert \pm 1\right\rangle $ states is also destroyed at $D_{c_{2}}$ and
then the system enters the large-$D$ phase with $\mathcal{O}^{12}=\mathcal{O}%
^{34}=0$.

\textit{Phase diagram.} Let us now explain our numerical results for the
evaluation of the phase diagram of the model. Our technique of choice has
been the so-called iTEBD algorithm \cite{Vidal-Orus-2007}. This algorithm
approximates the ground-state wave function of the system in the
thermodynamic limit by an MPS. To achieve this, the algorithm uses an
evolution in imaginary time. The parameter that controls the accuracy of the
approximation is the size of the matrices in the MPS approximation. This
parameter is usually called 'bond dimension', or $\chi $. Using this method,
we have computed MPS approximations to the ground state of the Hamiltonian
in Eq. (\ref{eq:Hamiltonian}) for different values of $D$ and $\chi $. Then,
for each one of these approximations we have computed the two SOPs given in
Eq. (\ref{eq:SOP}). In our simulations, we have seen that $\chi =40$ is
enough to reproduce the phase diagram of our model with sufficient accuracy
for our purposes.

Our results for the SOPs are shown in Fig. \ref{fig:SOPs}. We see clearly
that the two SOPs distinguish the three phases, exactly as predicted by the
effective field theory approach discussed above. In particular, we obtain
the values for the critical points $D_{c_{1}}\sim 0.08(1)$ and $%
D_{c_{2}}\sim 2.63(1)$. Interestingly, we see that the IH phase actually
extends through a large region in the phase diagram as compared to the SO(5)
Haldane phase.

\textit{Entanglement spectrum.} The concept of ES was introduced in Ref.
\cite{Haldane-2008} and has proven very useful in the characterization of 1D
gapped phases \cite{Pollman-2009}. From a mathematical point of view, the ES
is simply the spectrum of coefficients $\xi _{\alpha }=-2\log {\lambda
_{\alpha }}$, where $\lambda _{\alpha }$ are the Schmidt coefficients
obtained from the Schmidt decomposition of the ground state wave function
with respect to a bipartite partition in real space, $|\Psi \rangle
=\sum_{\alpha }\lambda _{\alpha }|\Psi _{\alpha }^{A}\rangle |\Psi _{\alpha
}^{B}\rangle $. In this equation $|\Psi _{\alpha }^{A(B)}\rangle $ are the
Schmidt vectors for the $A(B)$ subsystems. Quite importantly from a
numerical perspective, the iTEBD algorithm \cite{Vidal-Orus-2007}
automatically renders this information, since the MPS approximation to the
ground state wave function is always explicitly written in terms of the
coefficients $\lambda _{\alpha }$ for all possible bipartitions of the
system into two semi-infinite lines. Thus, the ES for these bipartitions can
be immediately read out from the numerical MPS wave function that
approximates the ground state.

\begin{figure}[tbp]
\centering
\includegraphics[scale=0.17]{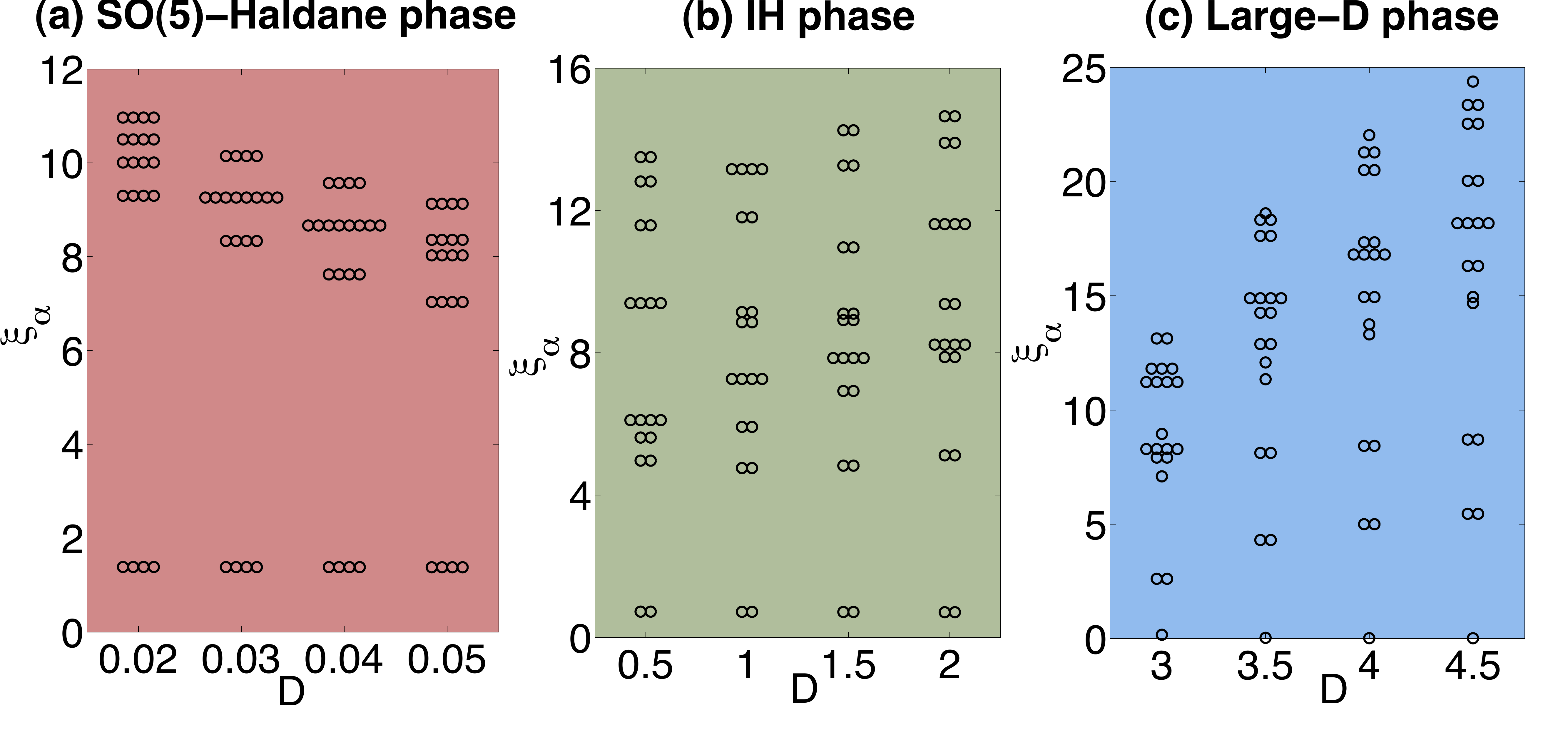}
\caption{(Color online) First 20 coefficients $\protect\xi_\protect\alpha$
of the entanglement spectrum for representative points in the (a) SO(5)
Haldane phase, (b) IH phase, and (c) large-$D$ phase.}
\label{fig:ES}
\end{figure}

In Fig. \ref{fig:ES} we show our results for the first 20 coefficients $%
\xi_{\alpha}$ of the ES at several representative points of the three
phases. As can be seen in the figure, the SO(5) Haldane phase is
characterized by a quadruple degeneracy in the ES (the coefficients organize
themselves in quadruplets), and the IH phase by a doubly degenerate ES (the
coefficients come in duplets). As expected, the degeneracy patterns in the
ES, introduced by a virtual cut, perfectly coincides with the physical edge
states in these two phases. Also, we see that the large-$D$ phase has no
characteristic degeneracy in the ES.

The robust degeneracies observed in the ES for the SO(5) Haldane and IH
phases are protected by the symmetries of the Hamiltonian. For the IH phase,
either (bond centered) spatial inversion or time reversal symmetry of Eq. (%
\ref{eq:Hamiltonian}) is sufficient to protect the doubly degenerate ES \cite%
{Pollman-2009}. However, the protection of the quadruply degenerate ES in
the SO(5) Haldane phase is beyond the scope of these two symmetries, and is
actually related to the $(Z_{2}\times Z_{2})^{2}$ symmetry. To prove this,
it is sufficient to show that $(Z_{2}\times Z_{2})^{2}$ allows a nontrivial
four-dimensional irreducible projective representation \cite%
{Gu-2009,Pollman-2009}. Let us identify such a projective representation by
focusing on the SO(5) point $D=0$. To simplify the notation, we switch from
Cartan basis to vector basis for SO(5). Using this notation, the five states
in the vector basis are written as $|n^{d}\rangle $ ($d=1,\ldots ,5$) and
the SO(5) generators are given by $L^{ab}=i(|n^{a}\rangle \langle n^{b}|-%
\mathrm{h.c.})$. The SO(5) point $D=0$ has a MPS ground state $|\Psi \rangle
=\sum_{\{a_{j}\}}\mathrm{Tr}(\Gamma ^{a_{1}}\lambda \cdots \Gamma
^{a_{N}}\lambda)|n^{a_{1}},\ldots ,n^{a_{N}}\rangle $, where $\lambda =\frac{%
1}{2}\mathds{1}_{4\times 4}$ is the (diagonal) matrix of Schmidt
coefficients, and the four-dimensional matrices $\Gamma ^{a}$ at each site
satisfy the Clifford algebra $\{\Gamma ^{a},\Gamma ^{b}\}=2\delta _{ab}$
\cite{Tu-2008}. Since $|\Psi \rangle $ is invariant under $(Z_{2}\times
Z_{2})^{2}$ rotations, the local $\Gamma $ matrices must satisfy the
transformation \cite{Cirac-2008}%
\begin{equation}
\sum_{d^{\prime }}(U^{ab})_{dd^{\prime }}\Gamma ^{d^{\prime }}=e^{i\theta
_{ab}}(V^{ab})^{\dagger }\Gamma ^{d}V^{ab}.  \label{eq:symmetry}
\end{equation}%
By using the Clifford algebra, we obtain $\theta _{ab}=0$ and the unitaries $%
V^{ab}=\frac{1}{2i}[\Gamma ^{a},\Gamma ^{b}]$, where $\{\mathds{1}%
,V^{12}\}\times \{\mathds{1},V^{15}\} \times \{\mathds{1} ,V^{34}\}\times \{%
\mathds{1},V^{35}\}$ is a four-dimensional irreducible projective
representation of $(Z_{2}\times Z_{2})^{2}$, satisfying $\{V^{12},V^{15}\}=%
\{V^{34},V^{35}\}=\{V^{15},V^{35}\} = 0$ and $[V^{12},V^{34}] =
[V^{12},V^{35}] =[V^{15},V^{34}] =0$. Since $[V^{ab},\lambda] = 0 \ \forall
\ V^{ab}$, the algebra of $V^{ab}$ guarantees a quadruply degenerate ES in
the SO(5) Haldane phase.

Moreover, the $(Z_{2}\times Z_{2})^{2}$ symmetry also contains a
two-dimensional irreducible projective representation, which protects the
doubly degenerate ES in the IH phase. This indicates that the $(Z_{2}\times
Z_{2})^{2}$ symmetry, besides (bond centered) spatial inversion or time
reversal symmetry, also protects the IH phase. Thus, we conclude that in the
presence of $(Z_{2}\times Z_{2})^{2}$ symmetry, the SO(5) Haldane and IH
phases are distinct symmetry-protected topological phases \cite{Gu-2009},
which must be separated by topological phase transitions (e.g., $D_{c_1}$ in
our model).

\textit{Conclusion.} Here we have studied a spin-2 quantum chain with an IH
phase. By combining effective field theory and numerical approaches we have
determined the phase diagram and the characteristic properties of the
phases, including edge states and SOPs. An analysis of the ES and the $%
(Z_{2}\times Z_{2})^{2}$ symmetry reveals that the IH and SO(5) Haldane
phases are distinct topological phases protected by symmetries. Our results are clear evidence of the existence
of an IH phase in quantum spin chains.

Compared to previous work \cite%
{Oshikawa-1992,Oshikawa-1995,Schollwock-1996,Nomura-1998,Hatsugai-2008,Tonegawa-2011}
looking for the IH phase by considering the spin-2 Heisenberg chain (or XXZ
chain) with uniaxial anisotropy, we emphasize that our starting point of the
model (\ref{eq:Hamiltonian}) is an SO(5) Haldane phase, which is quite
different from the conventional Haldane phase in the standard spin-2 Heisenberg
chain and Affleck-Kennedy-Lieb-Tasaki chain \cite{Zheng-2010,Jiang-2010}.
Nevertheless, the IH phase in our model satisfies all characteristic
features suggested by Oshikawa, \cite{Oshikawa-1992} and our approach has
the advantage that the guidance of a low-energy effective field theory\
and full characterization of phases under symmetries provide firm evidence
of the existence of an IH phase.

\textit{Acknowledgements.} H.H.T. acknowledges M. Cheng, H. Katsura, and
Z.-X. Liu for helpful discussions. R.O. acknowledges the EU.

\end{document}